\documentclass[journal]{IEEEtran}
\ifCLASSINFOpdf
  \usepackage[pdftex]{graphicx}
\else
\fi
\usepackage{lineno}

\usepackage[cmex10]{amsmath}
\usepackage{wasysym}

\begin{document}

\title{Approximation of radiative transfer for surface spectral features}

\author{Fr{\'e}d{\'e}ric~Schmidt
\thanks{F. Schmidt is with the GEOPS, Universit{\'e} Paris-Saclay, CNRS, GEOPS, 91405, Orsay, France, France e-mail: (see frederic.schmidt@univerite-paris-saclay.fr}
\thanks{F. Schmidt is with Institut Universitaire de France (IUF)}%
\thanks{Manuscript submitted October 21, 2022}}

\markboth{Submitted to IEEE Geo. Rem. Sens. Letters}%
{Schmidt: Approximation of radiative transfer}

\maketitle


\begin{abstract}
Remote sensing hyperspectral and more generally spectral instruments are common tools to decipher surface features in Earth and Planetary science. While linear mixture is the most common approximation for compounds detection (mineral, water, ice, etc...), the transfer of light in surface and atmospheric medium are highly non-linear. The exact simulation of non-linearities can be estimated at very high numerical cost. Here I propose a very simple non-linear form (that includes the regular linear area mixture) of radiative transfer to approximate surface spectral feature. I demonstrate that this analytical form is able to approximate the grain size and intimate mixture dependence of surface features. In addition, the same analytical form can approximate the effect of Martian mineral aerosols. Unfortunately, Earth aerosols are more complex (water droplet, water ice, soot,...) and are not expected to follow the same trend.
\end{abstract}


%
\IEEEpeerreviewmaketitle

\section{Introduction} 

\IEEEPARstart{R}{remote} sensing hyperspectral images and spectral data are highly valuable to decipher surface's characteristics, such as composition but also grain size, roughness... The standard approximation for data analysis is the linear mixing. This approximation has been used to propose algorithms to estimate abundances, knowing the endmembers (usually laboratory of pure chemical compounds), incorporating several level of complexity, such as positivity constraint~\cite{Heinz_FullyConstrainedLinear_TGRS2001}, sparsity \cite{Iordache_SparseUnmixing_TGRS2011}, endmember variability \cite{Drumetz_SpectralUnmixingDerivation_IGaRSL2020}... Also this approximation can be used in a non-supervised manner, leading to other class of algorithm such as principal/independent component analysis \cite{Cardoso_JADE_IEEEsigProcWork1993}, blind source separation \cite{Moussaoui_JADE-BPSS_Neurocomp2008},... This type of approach has been reviewed in \cite{Bioucas-Dias_HyperspectralUnmixingOverview_JSTAR2012}. Adding non-linear complexity is often out of scope of such algorithms due to overwhelm complexity. More recently, some non-linearities have been included in the data analysis in bilinear form \cite{Nascimento_Nonlinearmixturemodel_SPIE2009}, using kernels \cite{Iordache_SparseUnmixing_TGRS2011}... This type of model has been reviewed in \cite{Heylen_ReviewNonlinearHyperspectral_IJSTARS2014, Dobigeon_NonlinearUnmixingHyperspectral_SPMI2014}. Unfortunately, the full non-linearities are only tractable using bayesian Monte Carlo inversion approach such as in \cite{Schmidt_RealisticuncertaintiesHapke_I2015, Andrieu_Icestateevolution_I2018, CruzMermy_MicrophysicEuropasSurface_2022}. The latter class of method is very powerful but is impossible to apply on large spectral data, especially when the endmembers are not known beforehand. The standard data analysis is thus two step : first, performing a relatively fast algorithm to identify the compounds, second a robust algorithm to quantify the surface properties.

In the spirit of \cite{Drumetz_SpectralUnmixingDerivation_IGaRSL2020} who proposed a simplification of Hapke's theory \cite{Hapke_Book1993}, here I propose a new approximation of radiative transfer in soil and martian atmosphere in order to mimic non-linear effect of light propagation in complex media. This formulation shall be used to define a new generation of algorithms.Please note that the Hapke's work \cite{Hapke_Book1993} and Shkuratov model \cite{Shkuratov_ModelReflectance_Icarus1999} are only valid in geometrical optic approximation, i.e. : for grain size larger than wavelength.

\subsection{Method} 

The standard mixing model used for spectra data analysis is the linear mixture of spectra :

\begin{equation}
\mathbf{Y} = \sum_{i=1}^{N} A_{i}.\mathbf{S_{i}} \label{eq:linear_mixing}
\end{equation}

With $\mathbf{Y}$ the simulation spectra, $\mathbf{S_{i}}$ the collection $N$ spectra, referred as spectral database,  $A_{i}$ the mixing coefficient, 
and $N_{j}$ the number of spectral bands. 

In order to easily mimic the effect of the main non-linear effect: grain size, intimate mixture and aerosols, I propose to add an alteration in form of:

\begin{equation}
\mathbf{Y} = \sum_{i=1}^{N} A_{i}.\mathbf{S_{i}}^\alpha + \beta \label{eq:linear_mixing_alteration}
\end{equation}

with $\alpha>0$ an exponent that model the change in band depth and $\beta$ a constant that model the brightness level change. $\mathbf{S_{i}}^\alpha$ do not represents a matrix exponentiation but a single value operation on each element of $\mathbf{S_{i}}$. If the grain size/aerosols are not homogeneous in the scene, $\alpha$ and $\beta$ may change spatially.

In order to validate the ability of this simple model to reproduce non-linear effect, we compare it with non-linear simulation such as in \cite{Schmidt_Mineralsdetectionhyperspectral_I2014}. The effect of grain size is simulated using the Shkuratov model \cite{Shkuratov_ModelReflectance_Icarus1999}, assuming a refractive index of 1.7, a porosity of 20 \% and a grain size $\phi$. We first estimate the imaginary part of the optical constant $\kappa$ and then regenerate a synthetic spectra with a new grain size $\phi * fact$. The grain size factor $fact$ varies from $10^{-3}$ to $10^{3}$. In case of granular mixture, we tested proportion of 100\%, 99.9\%, 99\%, 90\%, 66\%, 50\%  and their complementary part (called $prop$). We used the same 26 mineral spectra as in \cite{Schmidt_Mineralsdetectionhyperspectral_I2014}.

The effect of aerosols is simulated using a parametrization of optical thickness and single scattering albedo \cite{Vincendon_DustAerosols_JGR2007} using DISORT to solve the radiative transfer equation \cite{Stamnes_DISORT_AO1988, Lin_Improveddiscreteordinate_JoQSaRT2015}. We tested 8 Aerosols Optical Thickness (AOT) at 1 $\mu$m from 0.01 to 20 in log space.


A simple Nelder-Mead simplex minimization \cite{Nelder_SimplexMethodFunction_TCJ1965, Lagarias_ConvergencePropertiesNelder_SJoO1998} has been implemented to estimate $A$ (only in the case of mixture), $\alpha$ and $\beta$ that minimizes the RMS between the non-linear radiative transfer $\mathbf{Y_{RT}}$ and the approximation $\mathbf{Y}$ spectra :

\begin{equation}
RMS = \sqrt{ \frac{1}{N_{j}}   \sum ( \mathbf{Y}_{RT}-\mathbf{Y})^2   }   \label{eq:RMS}
\end{equation}

\subsection{Results} 

Except for few spectra with significant saturation, for all experiment using 26 different minerals with all $fact$, granular mixture, and aerosols scattering demonstrates, we found RMS $\apprle 10^{-2}$, justifying the approximation. 

\subsubsection{Grain size and granular mixture}

\begin{figure} 

\centering{}
\includegraphics[viewport=50bp 245bp 540bp 640bp,clip,width=1.0\columnwidth]{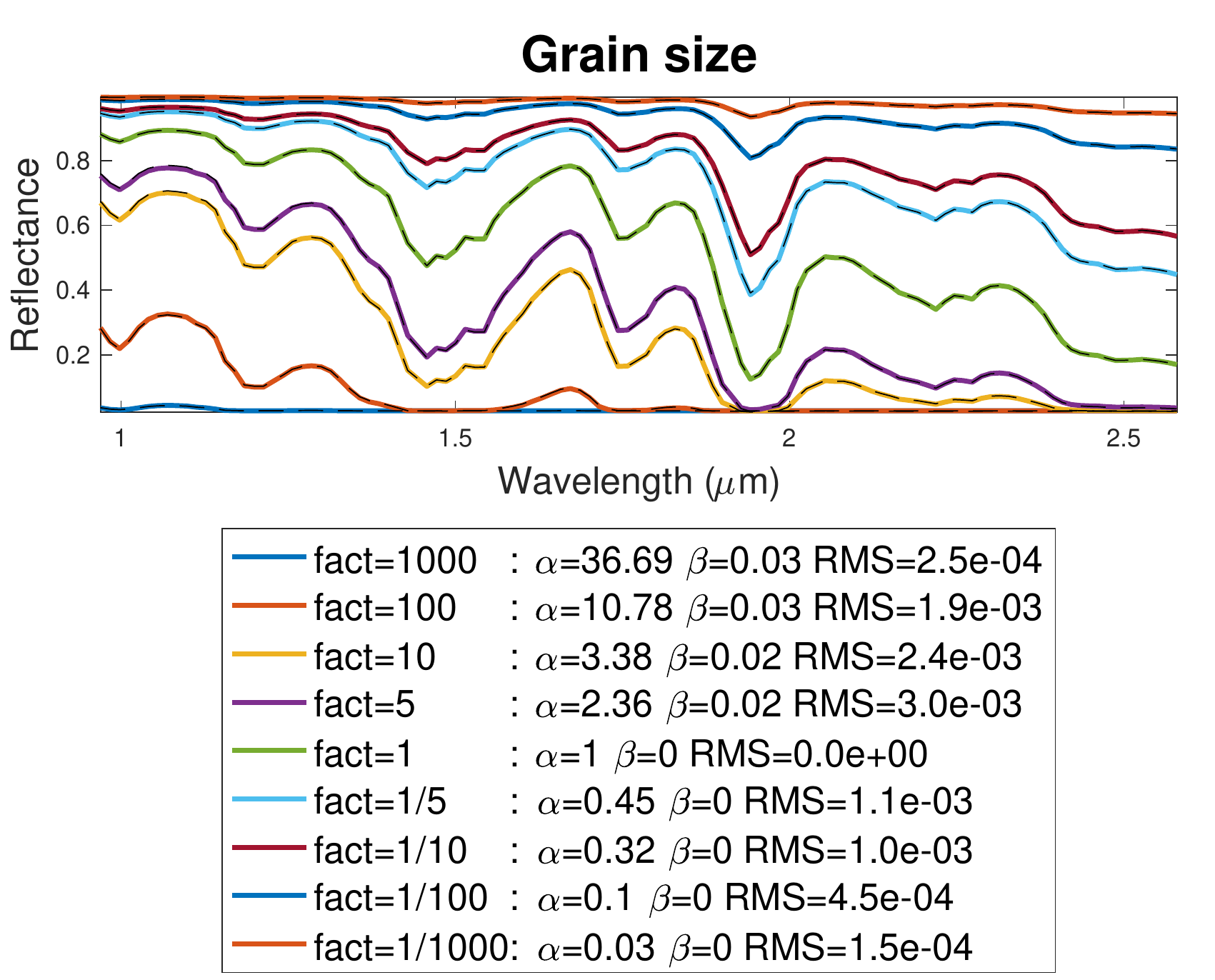}

\caption{Validation of the approximation of non-linear radiative transfer model for grain size effect on pure gypsum. Parameter $fact$ indicates the grain size factor change, for instance 1000 means that the grain size has been multiplied by 1000 respect to the reference spectrum.
\label{fig:Non-Linear-Simulation-Validation-GrainSize}}
\end{figure}

Figure \ref{fig:Non-Linear-Simulation-Validation-GrainSize} shows the cases for non-linear granular radiative transfer simulation and approximation using Eq. \ref{eq:linear_mixing_alteration} for a gypsum spectra as an example.   
A value of $\alpha <1$ is coherent with absorption bands that are weaker and occurs for smaller grain size. Whereas, a value of $\alpha >1$ is coherent with absorption bands that are stronger, toward saturation and occurs for larger grain size. The level correction $\beta =0$ for smaller grain size, but $\beta$ is positive and increases with larger grain size.

\begin{figure} 

\centering{}
\includegraphics[viewport=38bp 157bp 550bp 717bp, clip, width=1.0\columnwidth]{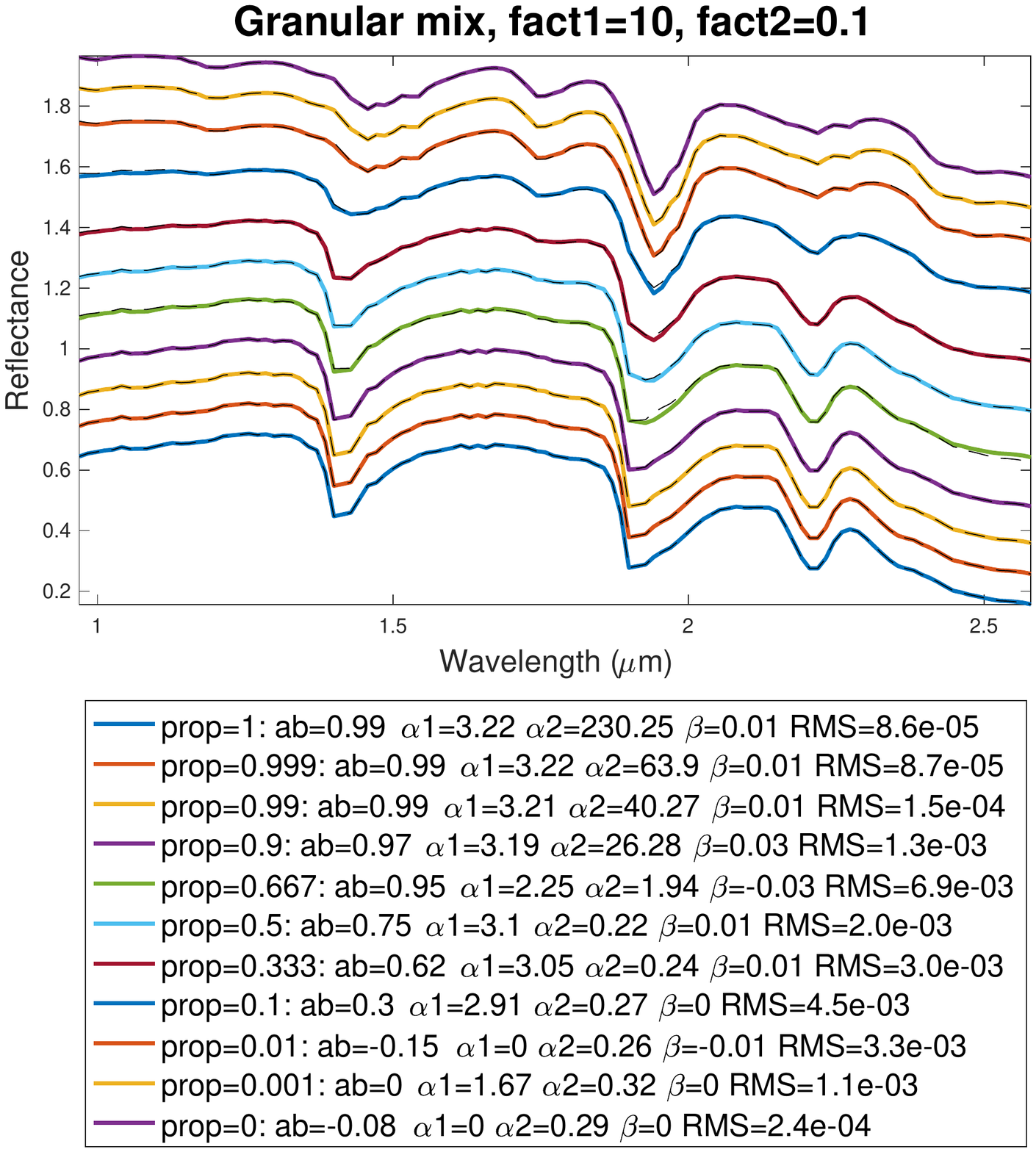}

\caption{Validation of the approximation of non-linear radiative transfer model for granular mixture of smectite (endmember 1) and gypsum (endmember 2). $prop$ is the imposed proportion and $ab$ is the retrieved abundance using the approximation. An offset of 0.1 is added for visibility.
\label{fig:Non-Linear-Simulation-Validation-GrainSize-granular-mixture}}
\end{figure}

Figure \ref{fig:Non-Linear-Simulation-Validation-GrainSize-granular-mixture} shows an example of granular mixture of gypsum and smectite. Again the spectral trend is well-fitted in this approximation (see for instance the 1.9 $\mu$m absorption band shape change). The abundances retrieved could significantly differ from the actual one but the trend is coherent. Coefficients $\alpha$ are also coherent with precedent finding. The level correction $\beta$ depends on relative brightness of both pure spectrum.

\begin{figure} 

\centering{}
\includegraphics[viewport=38bp 210bp 550bp 625bp,clip,width=1.0\columnwidth]{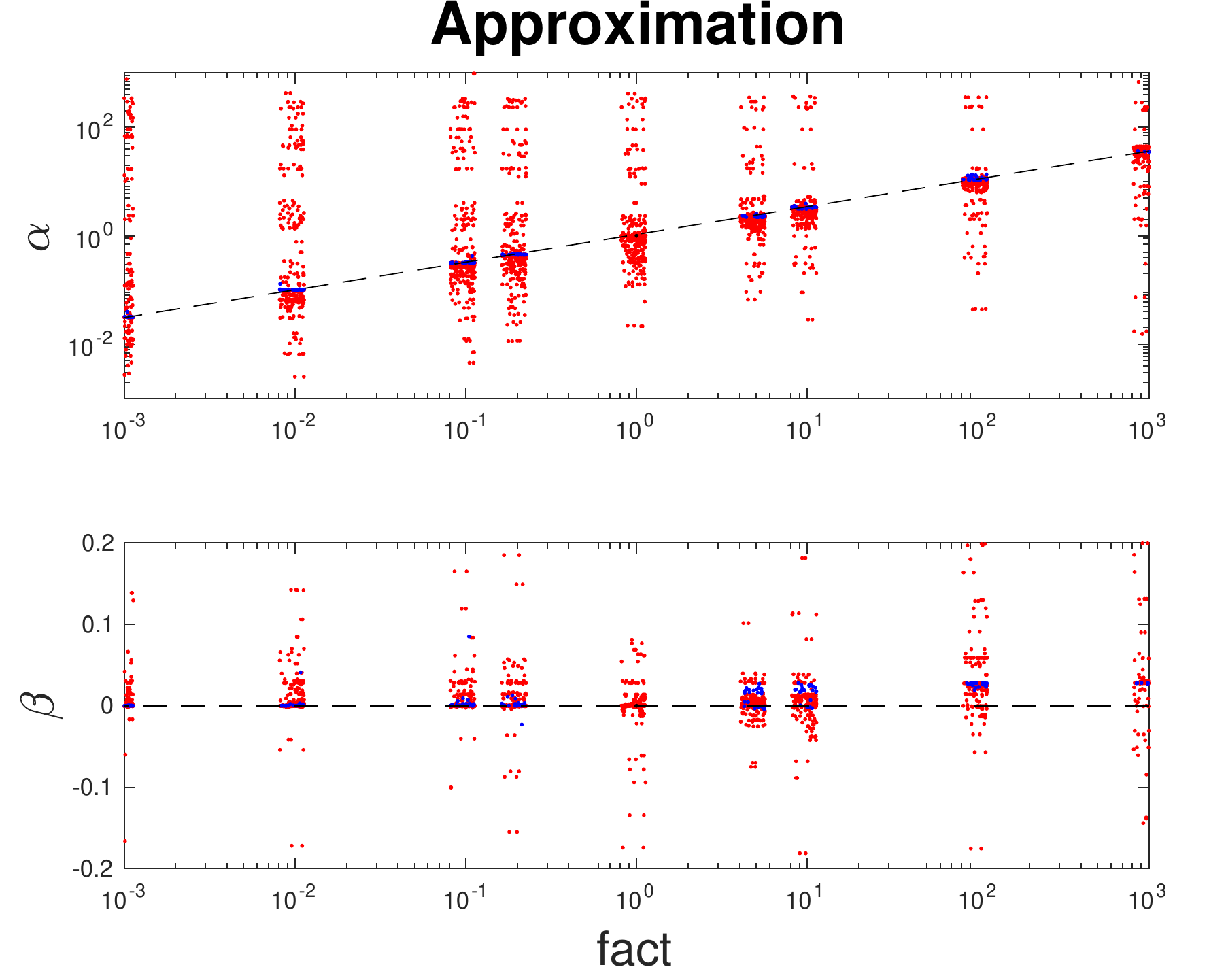}

\caption{Evaluation of $\alpha$ and $\beta$ as a function of grain size factor $fact$ compiling all results for: (in blue) grain size change of pure material. (in red) alteration of granular mixture with grain size change. The linear trend on blue points is $\log10(\alpha) = 0.5105 \times \log10(fact) + 0.0249$. The pearson correlation coefficient is 0.99957 for 179 points. In order to increase the visibility of each individual point, a random offset is added along the x axis just for plotting.
\label{fig:Non-Linear-Simulation-alpha-beta}}
\end{figure}

Figure \ref{fig:Non-Linear-Simulation-alpha-beta} summarizes the results of all experiments. The trends from Fig. \ref{fig:Non-Linear-Simulation-Validation-GrainSize} seem valid for all the 26 minerals when considering grain size change only (in blue). Unfortunately, the non-linearities are more complex in a case of granular mixture and there is no general trend neither for $\alpha$, nor $\beta$. Nevertheless, the RMS is lower than $10^{-2}$ for 98\% of the cases (maximum $4.1\times10^{-2}$) demonstrating that the approximation is valid in a very large range of minerals, grain size and granular mixture.

\subsubsection{Martian aerosols}

\begin{figure}

\centering{}
\includegraphics[viewport=50bp 245bp 540bp 640bp,clip,width=1.0\columnwidth]{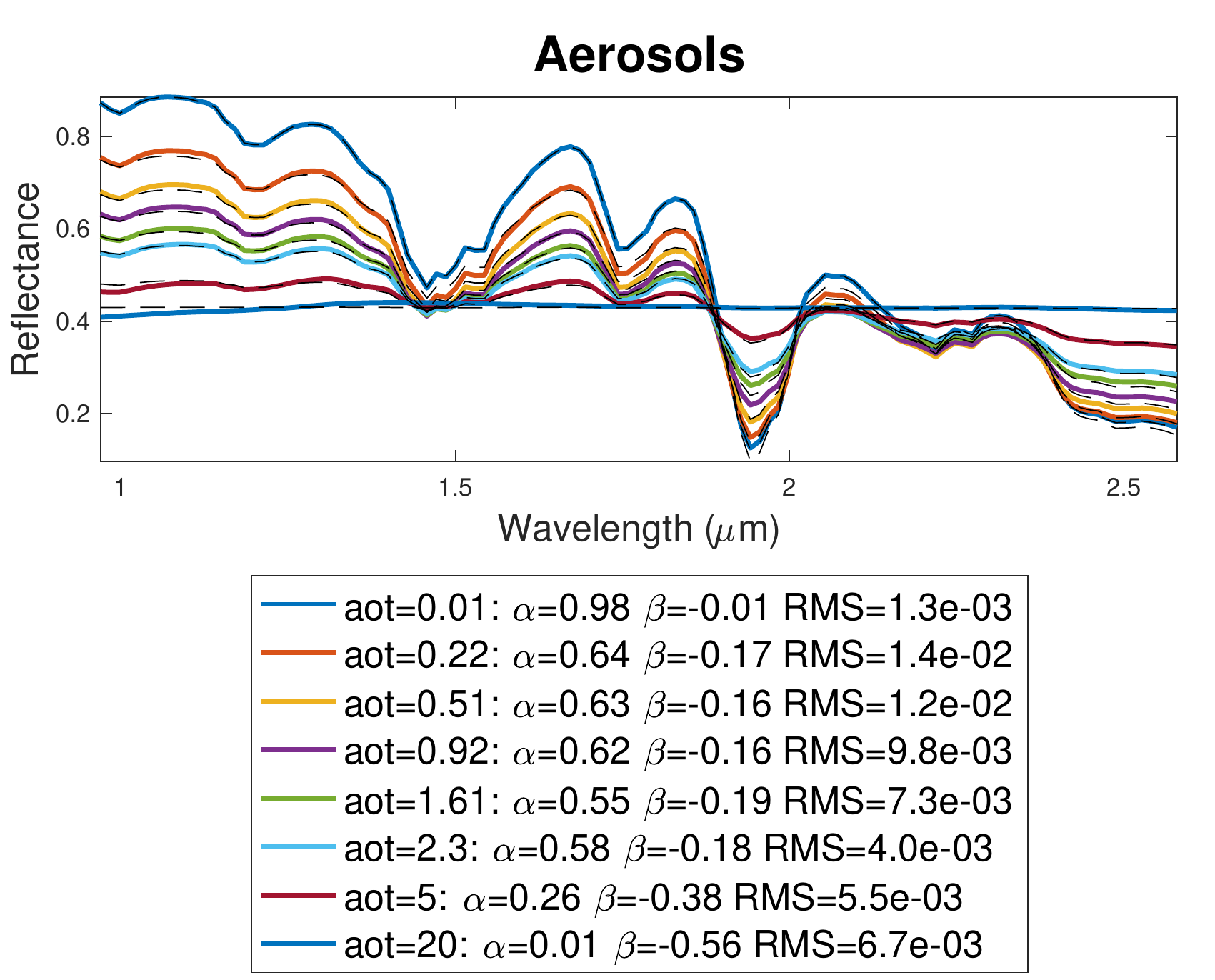}

\caption{Validation of the linear approximation of non-linear radiative transfer model on Martian aerosols effect on a gypsum spectrum. $aot$ stands for Aerosols Optical Thickness ($\tau$) and represents the aerosols quantity in the atmosphere with an attenuation of $e^{-\tau}$ at 1 microns.
\label{fig:Non-Linear-Simulation-Validation-Aerosols}}
\end{figure}

Fig. \ref{fig:Non-Linear-Simulation-Validation-Aerosols} shows the cases for non-linear aerosol scattering radiative transfer simulation and approximation using Eq. \ref{eq:linear_mixing_alteration} for soil made of pure gypsum as an example. A value of $\alpha <1$ is coherent with absorption bands that are weaker and occurs for aerosols alteration. The level correction $\beta$ is negative and decreases with larger AOT as expected since the pure spectra is brighter than the aerosols contribution. Again the RMS is lower than $10^{-2}$ for 96\% of the cases (maximum at $3.6\times10^{-2}$) demonstrating that the approximation is valid in a very large range of surface material and aerosols content for Mars. 

Please note that contrarily to Earth, Martian atmospheric gas absorption can be corrected independently since the aerosols are confined in the lower layers of the atmosphere \cite{Doute_PSPC_PSS2007}. Applying this strategy on Earth is unfortunately more difficult also because of the high variability of terrestrial aerosols (water ice, droplet, soot,...). As a perspective, I could propose to include the spectra of optically thick aerosols in the spectra endmember library.

\section{Conclusion} 

A new non-linear formulation is proposed to approximate radiative transfer in the surface granular material and in Martian aerosols scattering. Using simple fitting procedure, I demonstrated that this formulation is coherent with the full non-linear formulation and by construction to linear aerial mixture. Future algorithms should be based on this new formulation. Ideally, the methods should be able to estimate  $\alpha$ and $\beta$, at the same time as $A$ (and $\mathbf{S}$ in case of non-supervised approach). An application to real Martian data has been performed \cite{Gurgurewicz_MegashearshydrothermalismMartian_NCEE2022}. In a case of supervised unmixing, one could build an extended dictionary from the database of endmember spectra $S$ by applying several $\alpha$, using $\mathbf{S'}=\mathbf{S}^\alpha$. By adapting usual method using sparsity constraint, one should be able to estimate the abundances $A$.


%



\section*{Acknowledgment}
The author would like to thank support from the ``Institut National des Sciences de l'Univers'' (INSU), the ``Centre National de la Recherche Scientifique'' (CNRS) and ``Centre National d'Etudes Spatiales'' (CNES) through the ``Programme National de Plan\'etologie''.

\ifCLASSOPTIONcaptionsoff
  \newpage
\fi



\bibliographystyle{IEEEtran}
\bibliography{IEEEexample}

\begin{thebibliography}{10}
\providecommand{\url}[1]{#1}
\csname url@samestyle\endcsname
\providecommand{\newblock}{\relax}
\providecommand{\bibinfo}[2]{#2}
\providecommand{\BIBentrySTDinterwordspacing}{\spaceskip=0pt\relax}
\providecommand{\BIBentryALTinterwordstretchfactor}{4}
\providecommand{\BIBentryALTinterwordspacing}{\spaceskip=\fontdimen2\font plus
\BIBentryALTinterwordstretchfactor\fontdimen3\font minus
  \fontdimen4\font\relax}
\providecommand{\BIBforeignlanguage}[2]{{%
\expandafter\ifx\csname l@#1\endcsname\relax
\typeout{** WARNING: IEEEtran.bst: No hyphenation pattern has been}%
\typeout{** loaded for the language `#1'. Using the pattern for}%
\typeout{** the default language instead.}%
\else
\language=\csname l@#1\endcsname
\fi
#2}}
\providecommand{\BIBdecl}{\relax}
\BIBdecl

\bibitem{Heinz_FullyConstrainedLinear_TGRS2001}
D.~Heinz and Chein-I-Chang, ``Fully constrained least squares linear spectral
  mixture analysis method for material quantification in hyperspectral
  imagery,'' \emph{Geoscience and Remote Sensing, IEEE Transactions on},
  vol.~39, no.~3, pp. 529--545, 2001.

\bibitem{Iordache_SparseUnmixing_TGRS2011}
M.-D. Iordache, J.~Bioucas-Dias, and A.~Plaza, ``Sparse unmixing of
  hyperspectral data,'' \emph{Geoscience and Remote Sensing, IEEE
  Transactions}, vol.~49, no.~6, pp. 2014--2039, 2011.

\bibitem{Drumetz_SpectralUnmixingDerivation_IGaRSL2020}
L.~Drumetz, J.~Chanussot, and C.~Jutten, ``Spectral unmixing: A derivation of
  the extended linear mixing model from the hapke model,'' \emph{{IEEE}
  Geoscience and Remote Sensing Letters}, vol.~17, no.~11, pp. 1866--1870, nov
  2020.

\bibitem{Cardoso_JADE_IEEEsigProcWork1993}
J.~Cardoso and A.~Souloumiac, ``An efficient technique for the blind separation
  of complex sources,'' in \emph{Higher-Order Statistics, 1993., IEEE Signal
  Processing Workshop on}, 1993, pp. 275--279.

\bibitem{Moussaoui_JADE-BPSS_Neurocomp2008}
S.~Moussaoui, H.~Hauksd{\'o}ttir, F.~Schmidt, C.~Jutten, J.~Chanussot, D.~Brie,
  S.~Dout{\'e}, and J.~Benediktsson, ``On the decomposition of mars
  hyperspectral data by ica and bayesian positive source separation,''
  \emph{Neurocomputing}, vol.~71, no. 10-12, pp. 2194--2208, Jun. 2008.

\bibitem{Bioucas-Dias_HyperspectralUnmixingOverview_JSTAR2012}
J.~Bioucas-Dias, A.~Plaza, N.~Dobigeon, M.~Parente, Q.~Du, P.~Gader, and
  J.~Chanussot, ``Hyperspectral unmixing overview: Geometrical, statistical,
  and sparse regression-based approaches,'' \emph{Selected Topics in Applied
  Earth Observations and Remote Sensing, IEEE Journal of}, vol.~5, no.~2, pp.
  354--379, 2012.

\bibitem{Nascimento_Nonlinearmixturemodel_SPIE2009}
J.~M.~P. Nascimento and J.~M. Bioucas-Dias, ``Nonlinear mixture model for
  hyperspectral unmixing,'' in \emph{{SPIE} Proceedings}, L.~Bruzzone,
  C.~Notarnicola, and F.~Posa, Eds.\hskip 1em plus 0.5em minus 0.4em\relax
  {SPIE}, sep 2009.

\bibitem{Heylen_ReviewNonlinearHyperspectral_IJSTARS2014}
R.~Heylen, M.~Parente, and P.~Gader, ``A review of nonlinear hyperspectral
  unmixing methods,'' \emph{{IEEE} Journal of Selected Topics in Applied Earth
  Observations and Remote Sensing}, vol.~7, no.~6, pp. 1844--1868, jun 2014.

\bibitem{Dobigeon_NonlinearUnmixingHyperspectral_SPMI2014}
N.~Dobigeon, J.-Y. Tourneret, C.~Richard, J.~Bermudez, S.~McLaughlin, and
  A.~Hero, ``Nonlinear unmixing of hyperspectral images: Models and
  algorithms,'' \emph{Signal Processing Magazine, IEEE}, vol.~31, no.~1, pp.
  82--94, Jan 2014.

\bibitem{Schmidt_RealisticuncertaintiesHapke_I2015}
F.~Schmidt and J.~Fernando, ``Realistic uncertainties on hapke model parameters
  from photometric measurement,'' \emph{Icarus}, vol. 260, pp. 73 -- 93, 2015.

\bibitem{Andrieu_Icestateevolution_I2018}
F.~Andrieu, F.~Schmidt, S.~Dout{\'{e}}, and E.~Chassefi{\`{e}}re, ``Ice state
  evolution during spring in richardson crater, mars,'' \emph{Icarus}, vol.
  315, pp. 158--173, nov 2018.

\bibitem{CruzMermy_MicrophysicEuropasSurface_2022}
G.~Cruz~Mermy, F.~Schmidt, F.~Andrieu, I.~Belgacem, T.~Cornet, and
  N.~Altobelli, ``Selection of chemical species for europa’s surface using
  galileo/nims,'' \emph{Icarus (under review)}, 2022.

\bibitem{Hapke_Book1993}
B.~{Hapke}, \emph{{Theory of reflectance and emittance spectroscopy}}.\hskip
  1em plus 0.5em minus 0.4em\relax Topics in Remote Sensing, Cambridge, UK:
  Cambridge University Press, 1993.

\bibitem{Shkuratov_ModelReflectance_Icarus1999}
Y.~Shkuratov, L.~Starukhina, H.~Hoffmann, and G.~Arnold, ``A model of spectral
  albedo of particulate surfaces: Implications for optical properties of the
  moon,'' \emph{Icarus}, vol. 137, no.~2, pp. 235--246, Feb. 1999.

\bibitem{Schmidt_Mineralsdetectionhyperspectral_I2014}
F.~Schmidt, M.~Legendre, and S.~Le~Mou{\"e}lic, ``Minerals detection for
  hyperspectral images using adapted linear unmixing: Linmin,'' \emph{Icarus},
  vol. 237, no.~0, pp. 61--74, Jul. 2014.

\bibitem{Vincendon_DustAerosols_JGR2007}
M.~{Vincendon}, Y.~{Langevin}, F.~{Poulet}, J.-P. {Bibring}, and B.~{Gondet},
  ``{Recovery of surface reflectance spectra and evaluation of the optical
  depth of aerosols in the near-IR using a Monte Carlo approach: Application to
  the OMEGA observations of high-latitude regions of Mars},'' \emph{Journal of
  Geophysical Research (Planets)}, vol. 112, pp. E08S13, 1--12, Jul. 2007.

\bibitem{Stamnes_DISORT_AO1988}
K.~{Stamnes}, S.-C. {Tsay}, K.~{Jayaweera}, and W.~{Wiscombe}, ``{Numerically
  stable algorithm for discrete-ordinate-method radiative transfer in multiple
  scattering and emitting layered media},'' \emph{Appl. Opt.}, vol.~27, pp.
  2502--2509, Jun. 1988.

\bibitem{Lin_Improveddiscreteordinate_JoQSaRT2015}
Z.~Lin, S.~Stamnes, Z.~Jin, I.~Laszlo, S.-C. Tsay, W.~Wiscombe, and K.~Stamnes,
  ``Improved discrete ordinate solutions in the presence of an anisotropically
  reflecting lower boundary: Upgrades of the \{DISORT\} computational tool,''
  \emph{Journal of Quantitative Spectroscopy and Radiative Transfer}, vol. 157,
  no.~0, pp. 119 -- 134, 2015.

\bibitem{Nelder_SimplexMethodFunction_TCJ1965}
J.~A. Nelder and R.~Mead, ``A simplex method for function minimization,''
  \emph{The Computer Journal}, vol.~7, no.~4, pp. 308--313, jan 1965.

\bibitem{Lagarias_ConvergencePropertiesNelder_SJoO1998}
J.~C. Lagarias, J.~A. Reeds, M.~H. Wright, and P.~E. Wright, ``Convergence
  properties of the nelder--mead simplex method in low dimensions,''
  \emph{{SIAM} Journal on Optimization}, vol.~9, no.~1, pp. 112--147, jan 1998.

\bibitem{Doute_PSPC_PSS2007}
S.~{Dout{\'e}}, B.~{Schmitt}, Y.~{Langevin}, J.-P. {Bibring}, F.~{Altieri},
  G.~{Bellucci}, B.~{Gondet}, F.~{Poulet}, and {the MEX OMEGA team}, ``{South
  Pole of Mars: Nature and composition of the icy terrains from Mars Express
  OMEGA observations},'' \emph{Planetary and Space Science}, vol.~55, pp.
  113--133, Jan. 2007.

\bibitem{Gurgurewicz_MegashearshydrothermalismMartian_NCEE2022}
J.~Gurgurewicz, D.~M{\`{e}}ge, F.~Schmidt, S.~Dout{\'{e}}, and B.~Langlais,
  ``Megashears and hydrothermalism at the martian crustal dichotomy in valles
  marineris,'' \emph{Nature Communications Earth {\&} Environment}, vol.~3,
  no.~1, nov 2022.

\end{thebibliography}

\end{document}